\documentclass[preprint,preprintnumbers,amsmath,amssymb,superscriptaddress]{revtex4-1}

\usepackage{graphicx}
\usepackage{dcolumn}
\usepackage{bm}

\begin{document}

\title{Magnetic neutron scattering on nanocomposites: decrypting cross-section images using micromagnetic simulations}

\author{Sergey Erokhin}
\author{Dmitry Berkov}
\author{Nataliya Gorn}
\affiliation{INNOVENT Technology Development, Jena, Germany}
\author{Andreas Michels}
\email[Corresponding author. Electronic address: ]{andreas.michels@uni.lu}
\affiliation{Laboratory for the Physics of Advanced Materials, University of Luxembourg, Luxembourg}

\date{\today}

\begin{abstract}
We have used numerical micromagnetics for the calculation of the magnetic (small-angle) neutron scattering cross section of nanocomposites. The novel aspect of our approach consists in the possibility to study the applied-field dependence of the \emph{individual contributions} to the total magnetic scattering. Such a micromagnetic tool ideally complements neutron experiments in which one generally measures only a \emph{weighted sum} of the Fourier components of the magnetization. The procedure furnishes unique and fundamental information regarding the magnetic microstructure and corresponding magnetic scattering from nanomagnets. In particular, our simulation results explain the recent observation of dipolar correlations in two-phase nanocomposites and provide an answer to the question of the explicit dependence of the magnetization Fourier coefficients on the scattering vector.
\end{abstract}


\maketitle\

{\it Introduction.}---Magnetic neutron scattering and, in particular, magnetic small-angle neutron scattering (SANS) is a very powerful technique for the investigation of spin structures in magnetic materials on a length scale between $\sim 1$~nm and a few hundred of nm (for recent reviews see, e.g., Refs.~\cite{kohlbrecher05,michels08rop,albi2010}). The unique feature of SANS is the possibility to study magnetic structure in the bulk of materials, in contrast to various spectroscopic techniques, which mostly provide information about the magnetization state at or near the sample surface. Magnetic SANS has been employed, for instance, to study the range of magnetic correlations in nanocrystalline 3$d$ transition metals \cite{loeff00prl,michels03prl}, the vortex lattice of type-II superconductors \cite{forgan06}, magnetization dynamics in ferrofluids \cite{albi06}, magnetic domains in Nd$_2$Fe$_{14}$B permanent magnets \cite{kreyssig07}, nanocrystalline Tb with random paramagnetic suceptibility \cite{weissm08}, the spin-helix chirality in FeCoSi single crystals \cite{gri09}, so-called skyrmions in MnSi \cite{pflei2009}, electric-field-induced magnetization in multiferroic HoMnO$_3$ \cite{uehland2010}, the spin structure of core-shell nanoparticles \cite{krycka2010}, or the impact of heterogeneities on the magnetostriction of FeGa alloys \cite{laver2010}.

The quantity of interest in a magnetic SANS experiment is the elastic magnetic differential scattering cross section $d \Sigma / d \Omega$, which is usually recorded on a two-dimensional position-sensitive detector. Basic scattering theory prescribes that $d \Sigma / d \Omega$ can be expressed in terms of the Fourier coefficient $\mathbf{\widetilde{M}} = \mathbf{\widetilde{M}}(\mathbf{q})$ of the magnetization vector $\mathbf{M}(\mathbf{x})$, more specifically, $d \Sigma / d \Omega$ is a weighted sum of the products of Cartesian components of $\mathbf{\widetilde{M}}$. For bulk ferromagnets, $\mathbf{\widetilde{M}}$ depends in a complicated manner on the momentum-transfer vector, applied magnetic field, and on the magnetic interaction parameters (exchange interaction, magnetic anisotropy, dipolar interaction), and only for special cases, e.g., in the approach-to-saturation regime, one can obtain approximate closed-form expressions for $\mathbf{\widetilde{M}}$ \cite{michels08rop}.

The fact that the experimental SANS pattern $d \Sigma / d \Omega$ is composed of several individual contributions often hampers the straightforward interpretation of recorded SANS data. While, in principle, some Fourier coefficients are accessible by the experiment, e.g., through the application of a saturating magnetic field or by exploiting the neutron-polarization degree of freedom via so-called SANSPOL or POLARIS methods (e.g., \cite{albi2010,krycka2010,michels2010epjb}), it is often difficult to unambiguously determine a particular scattering contribution without ``contamination'' by unwanted Fourier components. For instance, when the applied field is not large enough to completely saturate the sample, then the scattering along the field direction does not represent the pure nuclear SANS, but contains also the magnetic SANS due to the misaligned spins \cite{bischof07}.

In this Letter, we report the results of full-scale three-dimensional micromagnetic simulations of the magnetic SANS cross section of magnetic nanocomposites. Both numerical micromagnetics \cite{micmagref} and magnetic neutron scattering are well developed and established methods which are widely employed for studying magnetism in solid-state physics. As we will show in the following, it is their {\emph{combination}} which provides new insights into the fundamentals of magnetic SANS and, thus, into the magnetic microstructure of nanomagnets. In particular, the decisive advantage of this approach resides in the possibilitiy to study the contributions of the \emph{individual} Fourier components of the magnetization to $d \Sigma / d \Omega$---rather than their combination---and relate them to the underlying magnetic microstructure. This sheds new light on the ongoing discussion regarding the explicit momentum-transfer dependence of $d \Sigma / d \Omega$ \cite{dufour2011}. The micromagnetic computations have been adapted to the microstructure of the iron-based two-phase alloy NANO\-PERM for which experimental data exist \cite{michels06prb}.

{\it Details of the micromagnetic algorithm.}---In our micromagnetic model we have taken into account the four standard contributions to the total magnetic energy: external field, (uniaxial) magnetic anisotropy, exchange and dipolar interaction energies. The two-phase nanocomposite microstructure, consisting of magnetically ``hard'' iron-based particles embedded in a magnetically ``soft'' amorphous matrix, was generated by employing an algorithm described in Ref.~\cite{erokhin2011ieee}. The simulation volume (=~sample volume) is a rectangular box of size 125$\times$380$\times$380~nm$^3$, which was discretized into $N = 10^5$ mesh elements. The average size of a ``hard'' inclusion (nanocrystal) is $D = 10$~nm (as in NANO\-PERM \cite{michels06prb}), whereas the mesh size used to discretize the ``soft'' phase is two times smaller. This discretization scheme then limits the accessible range of momentum transfers (via the sampling theorem) to $q \lesssim q_{\mathrm{max}} \cong 1$~nm$^{-1}$. The volume fraction of the nanocrystallites is $x_C = 40$~$\%$, corresponding to about $8000$ nanocrystals in the simulation volume. Materials parameters for hard (``$h$'') and soft (``$s$'') phases are: magnetizations $M_h = 1750$~kA/m and $M_s = 550$~kA/m, anisotropy constants $K_h = 4.6 \times 10^4$~J/m$^3$ and $K_s = 1.0 \times 10^2$~J/m$^3$. As a value for the exchange-stiffness constant we used $A = 0.5 \times 10^{-11}$~J/m for interactions both within the soft phase and between the hard and soft phases. Equilibrium magnetization state of the system was found, as usual, by minimizing the total magnetic energy (for more details on our micromagnetic methodology see Ref.~\cite{erokhin2011ieee}). The computed SANS cross sections shown below represent averages over typically 8$-$16 independent random configurations of the hard crystallites.

{\it Magnetic SANS cross section.}---For the most commonly used scattering geometry in a magnetic SANS experiment, where the applied magnetic field $\mathbf{H} \parallel \mathbf{e}_z$ is perpendicular to the wave vector $\mathbf{k}_0 \parallel \mathbf{e}_x$ of the incident neutrons, the elastic magnetic SANS cross section $d \Sigma / d \Omega$  for unpolarized neutrons can be expressed as \cite{michels08rop}
\begin{eqnarray}
\label{sigmasans}
\frac{d \Sigma}{d \Omega}(\mathbf{q}) = \frac{8 \pi^3}{V} b_H^2 \left( |\widetilde{M}_x|^2 + |\widetilde{M}_y|^2 \cos^2\theta + |\widetilde{M}_z|^2 \sin^2\theta \right. \nonumber \\ \left. - (\widetilde{M}_y \widetilde{M}_z^{\ast} + \widetilde{M}_y^{\ast} \widetilde{M}_z) \sin\theta \cos\theta \right). \nonumber
\end{eqnarray}
$V$ is the scattering volume, $b_H = 2.699 \times 10^{-15}$~m/$\mu_{\mathrm{B}}$ ($\mu_{\mathrm{B}}$: Bohr magneton), $c^*$ is a quantity complex-conjugated to $c$, $\theta$ denotes the angle between the scattering vector $\mathbf{q}$ and $\mathbf{H}$, and $\widetilde{M}_{(x,y,z)}(\mathbf{q})$ are the Fourier transforms of the magnetization components $M_{(x,y,z)}(\mathbf{x})$. Note that in the small-angle limit and for this particular geometry $\mathbf{q} \cong q \, (0, \sin\theta, \cos\theta)$. Since the focus here is on magnetic spin-misalignment scattering, we have ignored the nuclear SANS.

\begin{figure}[tb]
\centering
\resizebox{1.0\columnwidth}{!}{\includegraphics{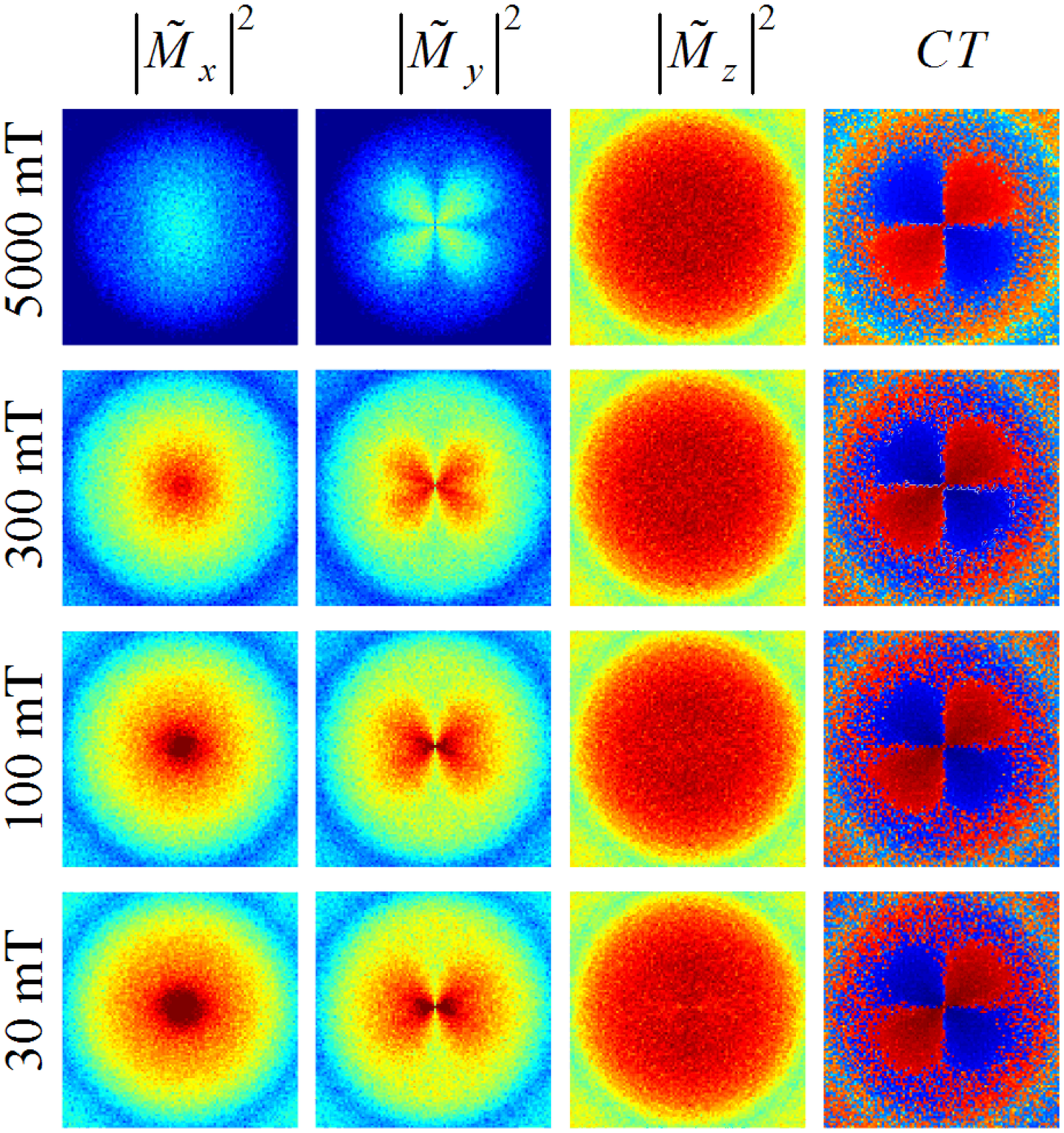}}
\caption{(color online). Results of the micromagnetic simulations for the Fourier coefficients of the magnetization. The images represent projections of the respective functions into the plane of the detector (i.e., $q_x = 0$). The external magnetic field $\mathbf{H} \parallel \mathbf{e}_z$ is applied horizontally in the plane of the detector. Values of $H$ decrease from top row (5~T) to bottom row (30~mT) (see insets). From left column to right column: $|\widetilde{M}_x|^2$, $|\widetilde{M}_y|^2$, $|\widetilde{M}_z|^2$, and $CT = - (\widetilde{M}_y \widetilde{M}_z^{\ast} + \widetilde{M}_y^{\ast} \widetilde{M}_z)$. Materials parameters of NANO\-PERM were used (see text). Pixels in the corners of the images have $q \cong 0.8$~nm$^{-1}$. Logarithmic color scale is used. In the first three columns from left, red color corresponds to ``high'' and blue color to ``low intensity''; in the fourth column, blue color corresponds to negative and red color to positive values of the $CT$.}
\label{fig1}
\end{figure}

\begin{figure}[tb]
\centering
\resizebox{1.0\columnwidth}{!}{\includegraphics{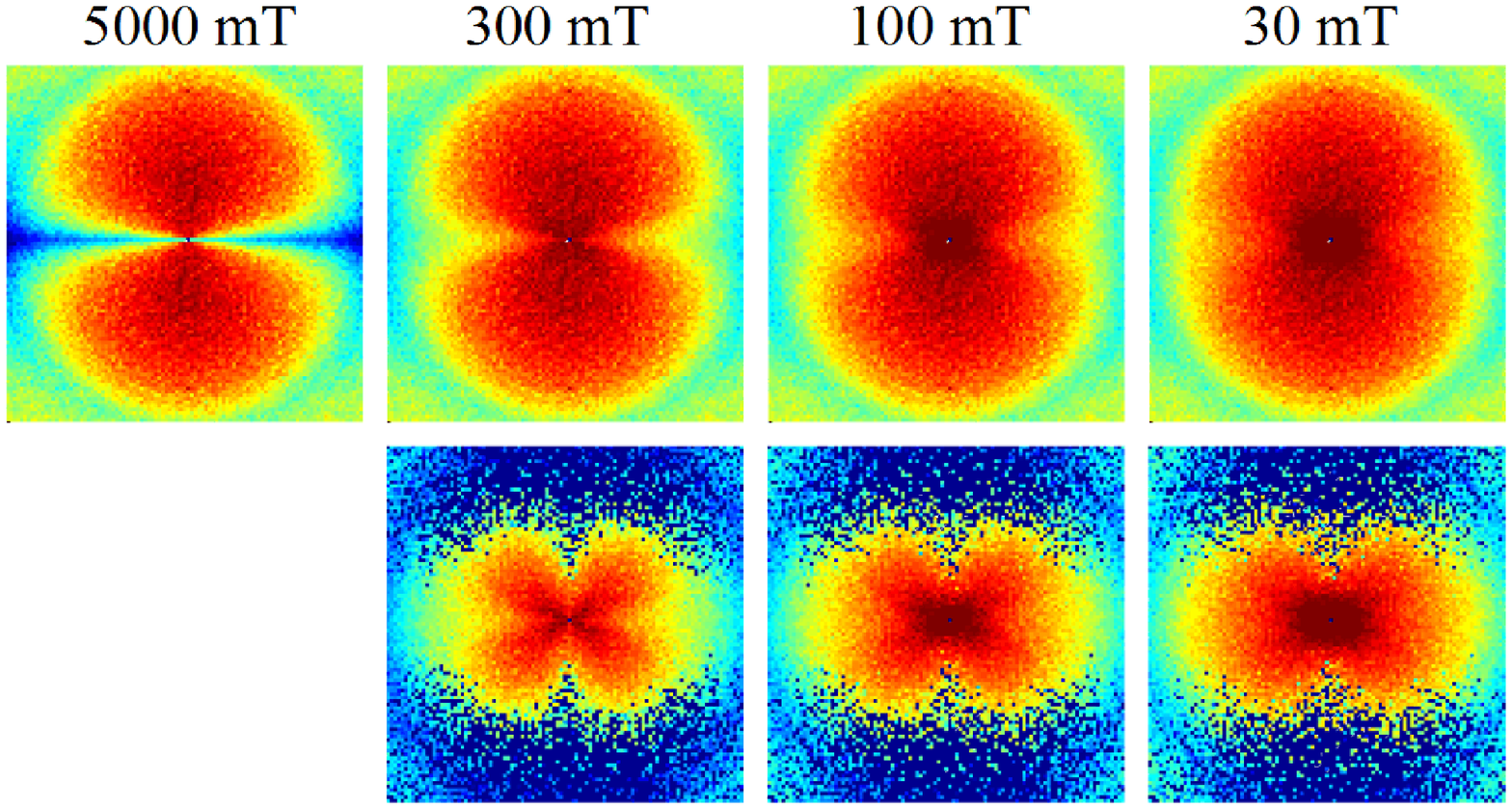}}
\caption{(color online). Applied-field dependence of the total magnetic SANS cross section $d \Sigma / d \Omega$ (upper row) and of the difference cross section (lower row) (where the SANS cross section at a saturating field of $\mu_0 H = 5$~T has been subtracted from the $d \Sigma / d \Omega$ at the respective field). Materials parameters and all other settings are as in Fig.~\ref{fig1}.}
\label{fig2}
\end{figure}

{\it Results and discussion.}---Figure~\ref{fig1} displays projections of the functions $|\widetilde{M}_x|^2$, $|\widetilde{M}_y|^2$, $|\widetilde{M}_z|^2$, and of the cross term $CT := - (\widetilde{M}_y \widetilde{M}_z^{\ast} + \widetilde{M}_y^{\ast} \widetilde{M}_z)$ into the plane of the two-dimensional detector at selected external-field values. Figure~\ref{fig2} shows the field dependence of the magnetic SANS cross section $d \Sigma / d \Omega$ (computed by means of the above expression for $d \Sigma / d \Omega$) and of the so-called difference cross section, where $d \Sigma / d \Omega$ at complete saturation (upper row left image in Fig.~\ref{fig2}) has been subtracted from the cross section at the respective field.
It can be seen in Fig.~\ref{fig1} that both $|\widetilde{M}_x|^2$ and $|\widetilde{M}_z|^2$ are isotropic (i.e., $\theta$ independent) over the whole field and $q$-range. By contrast, at the smallest $q$ and largest fields, the Fourier coefficient $|\widetilde{M}_y|^2$ reveals a pronounced angular anisotropy  with maxima roughly along the diagonals of the detector (the so-called ``clover-leaf'' anisotropy), whereas at the smaller fields, the anisotropy of $|\widetilde{M}_y|^2$ is rather of the $\cos^2\theta$-type (i.e., elongated parallel to $\mathbf{H}$). At saturation ($\mu_0 H = 5$~T), both $|\widetilde{M}_x|^2$ and $|\widetilde{M}_y|^2$ are relatively small and the main contribution to $d \Sigma / d \Omega$ is due the term $|\widetilde{M}_z|^2$, which originates from nanoscale jumps of the magnetization at phase boundaries. On decreasing the field, the transversal components increase in magnitude as long-range spin misalignment develops. The $CT$ oscillates in sign between quadrants on the detector; it is positive for $0^{\circ} < \theta < 90^{\circ}$, negative for $90^{\circ} < \theta < 180^{\circ}$, and so on. When the $CT$ is multiplied by $\sin\theta \cos\theta$, the corresponding contribution to the total $d \Sigma / d \Omega$ becomes positive-definite for all angles $\theta$. Therefore, and contrary to the common assumption that the $CT$ averages to zero for statistically isotropic polycrystalline microstructures, the $CT$ appears to be of special relevance in nanocomposite magnets.

\begin{figure}[tb]
\centering
\resizebox{0.50\columnwidth}{!}{\includegraphics{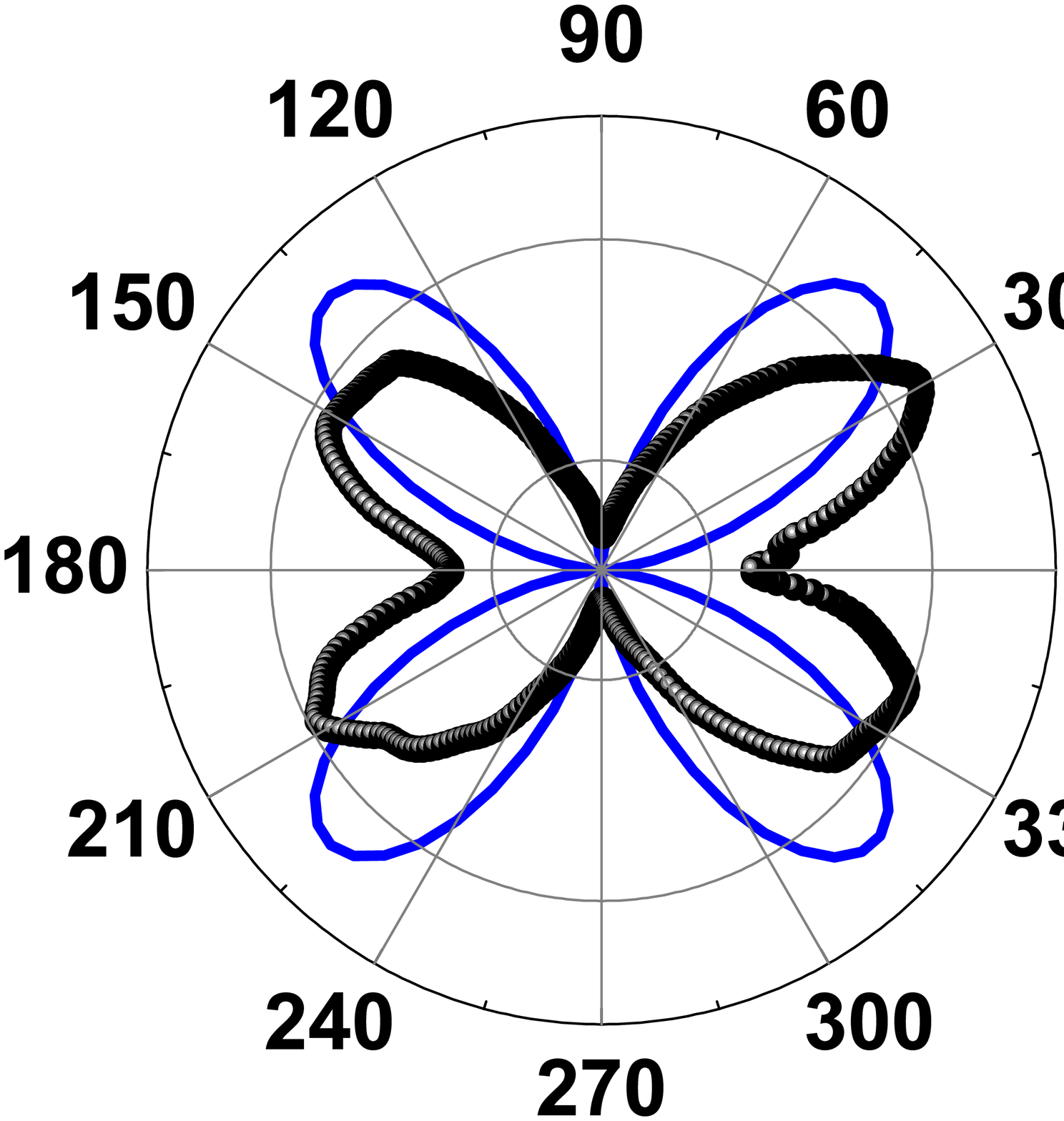}}
\caption{(color online). ($\bullet$) Polar plot of the Fourier coefficient $|\widetilde{M}_y|^2(\theta)$ at $q = (0.2 \pm 0.1)$~nm$^{-1}$ and $\mu_0 H = 0.3$~T. Data have been smoothed. Solid line: $|\widetilde{M}_y|^2 \propto \sin^2\theta \cos^2\theta$.}
\label{fig3}
\end{figure}

The finding that $|\widetilde{M}_x|^2$ and $|\widetilde{M}_z|^2$ are isotropic and that $|\widetilde{M}_y|^2 = |\widetilde{M}_y|^2(\theta)$ provides a straightforward explanation for the experimental observation of the clover-leaf anisotropy in the SANS data of the alloy NANO\-PERM \cite{michels06prb}. Our simulation results for the difference cross section $\propto (|\widetilde{M}_x|^2 + |\widetilde{M}_y|^2 \cos^2\theta + CT \sin\theta \cos\theta)$ (see Fig.~\ref{fig2}) agree qualitatively well with the experimental data \cite{erokhin2011ieee}. Note also that clover-leaf-type anisotropies in $d \Sigma / d \Omega$ have been reported for a number of other materials, including precipitates in steels \cite{bischof07}, nanocrystalline gadolinium \cite{michels08epl}, and nanoporous iron \cite{elmas09}. The maxima in $|\widetilde{M}_y|^2$ depend on $q$ and $H$, and on the magnetic interaction parameters and may appear at angles $\theta$ significantly smaller than 45$^{\circ}$, e.g., at $\theta \cong \pm 30^{\circ}$ (compare Fig.~\ref{fig3}).

\begin{figure}[tb]
\centering
\resizebox{0.60\columnwidth}{!}{\includegraphics{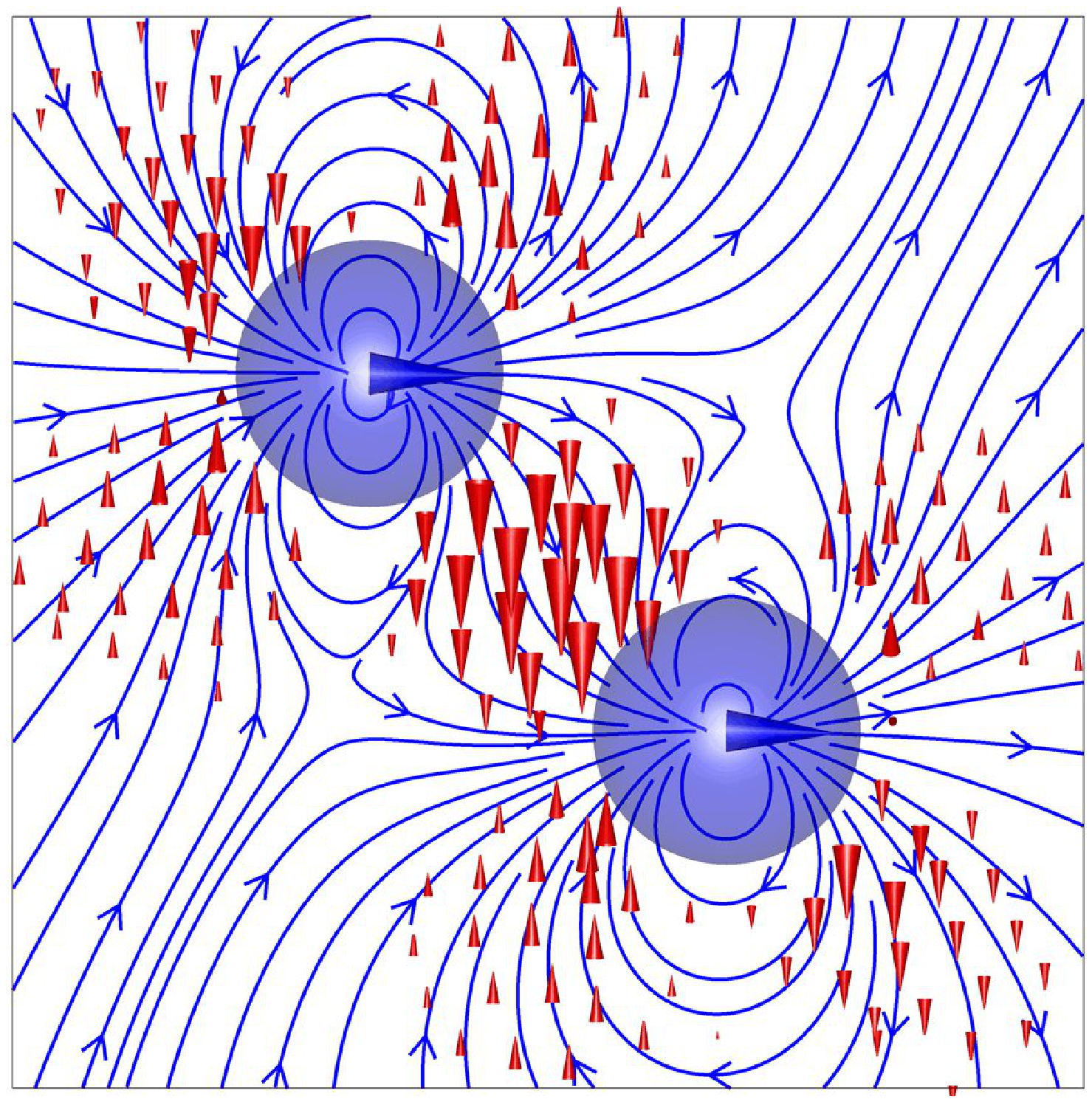}}
\caption{(color online). Two-dimensional real-space image of the computed spin distribution around two nanoparticles (violet circles). Red arrows: magnetization component $\mathbf{M}_\perp$ perpendicular to the applied field ($\mathbf{H}$ is horizontal in the plane); thickness of arrows is proportional to the magnitude of $\mathbf{M}_\perp$. Blue lines: dipolar field distribution.}
\label{fig4}
\end{figure}

As qualitatively discussed in Ref.~\cite{michels06prb}, the appearance of the clover-leaf anisotropy in $d \Sigma / d \Omega$ is related to the particular $\theta$ dependence of $\widetilde{M}_y$, which is imparted by virtue of the magnetodipolar interaction \cite{commentprl2012}. In fact, up to now, the physical origin for the existence of the clover-leaf in the magnetic SANS cross section was merely discussed in relation to the jump $\Delta M$ in the magnetization magnitude at the interface between the Fe particle and the amorphous magnetic matrix ($\Delta M \cong 1200$~kA/m for NANO\-PERM \cite{michels06prb}). This jump in magnetization gives rise to an inhomogeneous magnetodipolar field which decorates each nanoparticle and which causes nanoscale spin deviations within the matrix in the vicinity of each nanoparticle. As an illustration, Fig.~\ref{fig4} displays the real-space magnetization distribution around two nanoparticles. The symmetry of the spin structure replicates the symmetry of the $CT$ (compare to Fig.~\ref{fig1}). In the presence of an applied magnetic field the stray-field and associated magnetization configuration around each nanoparticle ``look'' similar (on the average), thus giving rise to \emph{dipolar correlations} which add up to a positive-definite $CT$ contribution to the magnetic $d \Sigma / d \Omega$.

\begin{figure}[tb]
\centering
\resizebox{1.0\columnwidth}{!}{\includegraphics{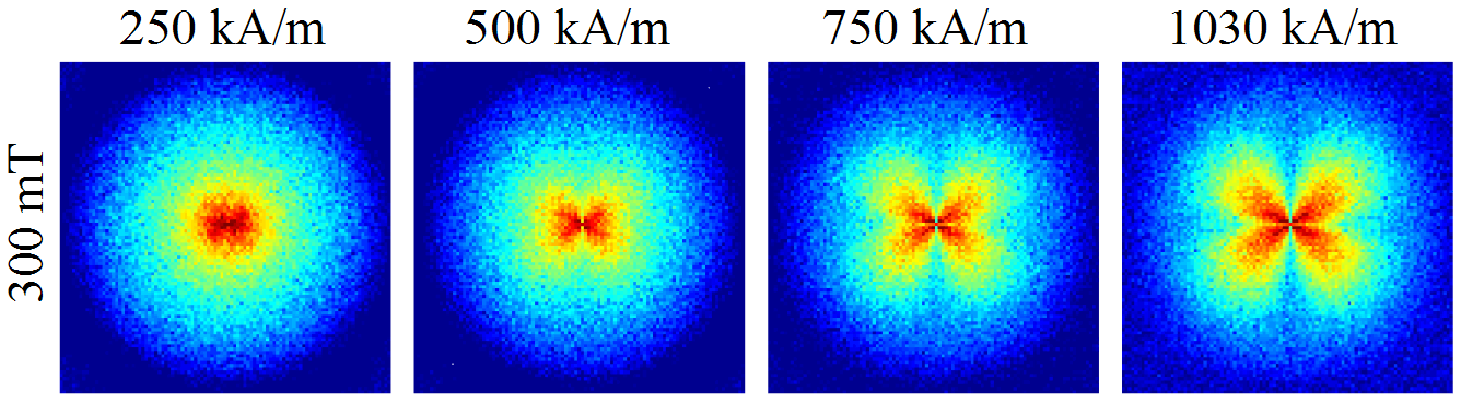}}
\caption{(color online). Fourier coefficient $|\widetilde{M}_y|^2$ at $\mu_0 H = 0.3$~T and for $M_h = M_s = M$ (i.e., $\Delta M = 0$). $M$ increases from left to right (see insets). $K_h = 4.6 \times 10^4$~J/m$^3$, $K_s = 1.0 \times 10^2$~J/m$^3$ and random variations in easy-axis directions from particle to particle. All other settings are as in Fig.~\ref{fig1}.}
\label{fig5}
\end{figure}

Next, we demonstrate that magnetodipolar correlations and the corresponding contribution of the $CT$ to the magnetic SANS cross section are of relevance for practically \emph{all} bulk magnetic materials which exhibit spatial variations in the magnetic parameters. In particular, not only variations in the magnetization magnitude (and possibly exchange coupling), but also variations in \emph{direction} and/or \emph{magnitude} of magnetic anisotropy $\bf{K}$ (random anisotropy) may give rise to corresponding dipolar correlations. In order to study the impact of such variations in $\bf{K}$ (which are, by construction, naturally included into our micromagnetic algorithm), we have computed the spin distribution for the situation that $M_h = M_s = M$ (i.e., $\Delta M = 0$) but for different values of $M$. The results for $|\widetilde{M}_y|^2$ are summarized in Fig.~\ref{fig5}. Note that $|\widetilde{M}_x|^2$ and $|\widetilde{M}_z|^2$ are both isotropic in this case (data not shown).

Figure~\ref{fig5} reveals that a clover-leaf-type pattern in $|\widetilde{M}_y|^2$ develops with increasing magnetization value $M$, i.e., with increasing strength of the magnetodipolar interaction. As jumps in $M$ at phase boundaries are excluded here as possible sources for perturbations in the spin structure, it is straightforward to conclude that nanoscale fluctuations in $\bf{K}$ give rise to inhomogeneous magnetization states (with $\nabla \cdot \mathbf{M} \neq 0$), which decorate each nanoparticle and which look similar to the structure shown in Fig.~\ref{fig4}. This observation strongly suggests that the origin of the clover-leaf pattern in $d \Sigma / d \Omega$ of nanomagnets is not only related to variations in magnetization magnitude but also due to variations in the magnitude and direction of the magnetic anisotropy field.

{\it Summary and conclusions.}---Using a recently developed micromagnetic simulation methodology we have computed the magnetic small-angle neutron scattering (SANS) cross section $d \Sigma / d \Omega$ of a two-phase nanocomposite. This approach allows one to study the applied-field dependence of the \emph{individual} scattering contributions to $d \Sigma / d \Omega$, in contrast to experiment, where generally only a \emph{weighted combination} of the magnetization Fourier coefficients is measured. It is this particular circumstance, in conjunction with the flexibility of our micromagnetic package in terms of microstructure variation (particle size and distribution, materials parameters, texture, etc.), which makes us believe that the approach of combining full-scale three-dimensional micromagnetic simulations with experimental magnetic-field-dependent SANS data will provide fundamental insights into the magnetic SANS of a wide range of magnetic materials. As we have demonstrated for the example of the iron-based two-phase alloy NANO\-PERM, we were able to explain on a deeper level the physical origin of the recently observed clover-leaf angular anisotropy in the magnetic SANS cross section. As a quite general result, our micromagnetic simulations suggest that magnetodipolar correlations (and the associated clover-leaf-shaped pattern in $d \Sigma / d \Omega$) are of relevance for all bulk magnets with inhomogeneous magnetic interaction parameters.

We thank Frank D\"obrich for critically reading the manuscript. This work was supported by the Deutsche Forschungsgemeinschaft (project No.~BE 2464/10-1) and by the National Research Fund of Luxembourg (ATTRACT project No.~FNR/A09/01 and project No.~FNR/10/AM2c/39).


\begin{thebibliography}{10}%
\makeatletter
\providecommand \@ifxundefined [1]{%
 \ifx #1\undefined \expandafter \@firstoftwo
 \else \expandafter \@secondoftwo
\fi
}%
\providecommand \@ifnum [1]{%
 \ifnum #1\expandafter \@firstoftwo
 \else \expandafter \@secondoftwo
\fi
}%
\providecommand \enquote [1]{``#1''}%
\providecommand \bibnamefont  [1]{#1}%
\providecommand \bibfnamefont [1]{#1}%
\providecommand \citenamefont [1]{#1}%
\providecommand\href[0]{\@sanitize\@href}%
\providecommand\@href[1]{\endgroup\@@startlink{#1}\endgroup\@@href}%
\providecommand\@@href[1]{#1\@@endlink}%
\providecommand \@sanitize [0]{\begingroup\catcode`\&12\catcode`\#12\relax}%
\@ifxundefined \pdfoutput {\@firstoftwo}{%
 \@ifnum{\z@=\pdfoutput}{\@firstoftwo}{\@secondoftwo}%
}{%
 \providecommand\@@startlink[1]{\leavevmode}%
 \providecommand\@@endlink[0]{}%
}{%
 \providecommand\@@startlink[1]{%
  \leavevmode
  \pdfstartlink
   attr{/Border[0 0 1 ]/H/I/C[0 1 1]}%
   user{/Subtype/Link/A<</Type/Action/S/URI/URI(#1)>>}%
  \relax
 }%
 \providecommand\@@endlink[0]{\pdfendlink}%
}%
\providecommand \url  [0]{\begingroup\@sanitize \@url }%
\providecommand \@url [1]{\endgroup\@href {#1}{\urlprefix}}%
\providecommand \urlprefix [0]{URL }%
\providecommand \Eprint[0]{\href }%
\@ifxundefined \urlstyle {%
  \providecommand \doi [1]{doi:\discretionary{}{}{}#1}%
}{%
  \providecommand \doi [0]{doi:\discretionary{}{}{}\begingroup
  \urlstyle{rm}\Url }%
}%
\providecommand \doibase [0]{http://dx.doi.org/}%
\providecommand \Doi[1]{\href{\doibase#1}}%
\providecommand \bibAnnote [3]{%
  \BibitemShut{#1}%
  \begin{quotation}\noindent
    \textsc{Key:}\ #2\\\textsc{Annotation:}\ #3%
  \end{quotation}%
}%
\providecommand \bibAnnoteFile [2]{%
  \IfFileExists{#2}{\bibAnnote {#1} {#2} {\input{#2}}}{}%
}%
\providecommand \typeout [0]{\immediate \write \m@ne }%
\providecommand \selectlanguage [0]{\@gobble}%
\providecommand \bibinfo [0]{\@secondoftwo}%
\providecommand \bibfield [0]{\@secondoftwo}%
\providecommand \translation [1]{[#1]}%
\providecommand \BibitemOpen[0]{}%
\providecommand \bibitemStop [0]{}%
\providecommand \bibitemNoStop [0]{.\EOS\space}%
\providecommand \EOS [0]{\spacefactor3000\relax}%
\providecommand \BibitemShut [1]{\csname bibitem#1\endcsname}%
\bibitem{kohlbrecher05}%
  \BibitemOpen
  \bibfield{author}{%
  \bibinfo {author} {\bibfnamefont{W.}~\bibnamefont{Wagner}}\ and\ \bibinfo
  {author} {\bibfnamefont{J.}~\bibnamefont{Kohlbrecher}},\ }%
  in\ \emph{\bibinfo {booktitle} {Modern Techniques for Characterizing Magnetic
  Materials}},\ \bibinfo {editor} {edited by\ \bibinfo {editor}
  {\bibfnamefont{Y.}~\bibnamefont{Zhu}}}\ (\bibinfo {publisher} {Kluwer
  Academic Publishers},\ \bibinfo {address} {Boston},\ \bibinfo {year} {2005})\
  pp.\ \bibinfo {pages} {65--103}%
  \bibAnnoteFile{NoStop}{kohlbrecher05}%
\bibitem{michels08rop}%
  \BibitemOpen
  \bibfield{author}{%
  \bibinfo {author} {\bibfnamefont{A.}~\bibnamefont{Michels}}\ and\ \bibinfo
  {author} {\bibfnamefont{J.}~\bibnamefont{Weissm\"uller}},\ }%
  \bibfield{journal}{%
  \bibinfo {journal} {Rep. Prog. Phys.}\ }%
  \textbf{\bibinfo {volume} {71}},\ \bibinfo {pages} {066501} (\bibinfo {year}
  {2008})%
  \bibAnnoteFile{NoStop}{michels08rop}%
\bibitem{albi2010}%
  \BibitemOpen
  \bibfield{author}{%
  \bibinfo {author} {\bibfnamefont{A.}~\bibnamefont{Wiedenmann}},\ }%
  \bibfield{journal}{%
  \bibinfo {journal} {Collection SFN}\ }%
  \textbf{\bibinfo {volume} {11}},\ \bibinfo {pages} {219} (\bibinfo {year}
  {2010}),\ \bibinfo {note} {http://www.neutron-sciences.org/}%
  \bibAnnoteFile{NoStop}{albi2010}%
\bibitem{loeff00prl}%
  \BibitemOpen
  \bibfield{author}{%
  \bibinfo {author} {\bibfnamefont{J.~F.}\ \bibnamefont{L\"{o}ffler}}, \bibinfo
  {author} {\bibfnamefont{H.~B.}\ \bibnamefont{Braun}},\ and\ \bibinfo {author}
  {\bibfnamefont{W.}~\bibnamefont{Wagner}},\ }%
  \bibfield{journal}{%
  \bibinfo {journal} {Phys. Rev. Lett.}\ }%
  \textbf{\bibinfo {volume} {85}},\ \bibinfo {pages} {1990} (\bibinfo {year}
  {2000})%
  \bibAnnoteFile{NoStop}{loeff00prl}%
\bibitem{michels03prl}%
  \BibitemOpen
  \bibfield{author}{%
  \bibinfo {author} {\bibfnamefont{A.}~\bibnamefont{Michels}}, \bibinfo
  {author} {\bibfnamefont{R.~N.}\ \bibnamefont{Viswanath}}, \bibinfo {author}
  {\bibfnamefont{J.~G.}\ \bibnamefont{Barker}}, \bibinfo {author}
  {\bibfnamefont{R.}~\bibnamefont{Birringer}},\ and\ \bibinfo {author}
  {\bibfnamefont{J.}~\bibnamefont{Weissm\"uller}},\ }%
  \bibfield{journal}{%
  \bibinfo {journal} {Phys. Rev. Lett.}\ }%
  \textbf{\bibinfo {volume} {91}},\ \bibinfo {pages} {267204} (\bibinfo {year}
  {2003})%
  \bibAnnoteFile{NoStop}{michels03prl}%
\bibitem{forgan06}%
  \BibitemOpen
  \bibfield{author}{%
  \bibinfo {author} {\bibfnamefont{M.}~\bibnamefont{Laver}}, \bibinfo {author}
  {\bibfnamefont{E.~M.}\ \bibnamefont{Forgan}}, \bibinfo {author}
  {\bibfnamefont{S.~P.}\ \bibnamefont{Brown}}, \bibinfo {author}
  {\bibfnamefont{D.}~\bibnamefont{Charalambous}}, \bibinfo {author}
  {\bibfnamefont{D.}~\bibnamefont{Fort}}, \bibinfo {author}
  {\bibfnamefont{C.}~\bibnamefont{Bowell}}, \bibinfo {author}
  {\bibfnamefont{S.}~\bibnamefont{Ramos}}, \bibinfo {author}
  {\bibfnamefont{R.~J.}\ \bibnamefont{Lycett}}, \bibinfo {author}
  {\bibfnamefont{D.~K.}\ \bibnamefont{Christen}}, \bibinfo {author}
  {\bibfnamefont{J.}~\bibnamefont{Kohlbrecher}}, \bibinfo {author}
  {\bibfnamefont{C.~D.}\ \bibnamefont{Dewhurst}},\ and\ \bibinfo {author}
  {\bibfnamefont{R.}~\bibnamefont{Cubitt}},\ }%
  \bibfield{journal}{%
  \bibinfo {journal} {Phys. Rev. Lett.}\ }%
  \textbf{\bibinfo {volume} {96}},\ \bibinfo {pages} {167002} (\bibinfo {year}
  {2006})%
  \bibAnnoteFile{NoStop}{forgan06}%
\bibitem{albi06}%
  \BibitemOpen
  \bibfield{author}{%
  \bibinfo {author} {\bibfnamefont{A.}~\bibnamefont{Wiedenmann}}, \bibinfo
  {author} {\bibfnamefont{U.}~\bibnamefont{Keiderling}}, \bibinfo {author}
  {\bibfnamefont{K.}~\bibnamefont{Habicht}}, \bibinfo {author}
  {\bibfnamefont{M.}~\bibnamefont{Russina}},\ and\ \bibinfo {author}
  {\bibfnamefont{R.}~\bibnamefont{G\"ahler}},\ }%
  \bibfield{journal}{%
  \bibinfo {journal} {Phys. Rev. Lett.}\ }%
  \textbf{\bibinfo {volume} {97}},\ \bibinfo {pages} {057202} (\bibinfo {year}
  {2006})%
  \bibAnnoteFile{NoStop}{albi06}%
\bibitem{kreyssig07}%
  \BibitemOpen
  \bibfield{author}{%
  \bibinfo {author} {\bibfnamefont{A.}~\bibnamefont{Kreyssig}}, \bibinfo
  {author} {\bibfnamefont{R.}~\bibnamefont{Prozorov}}, \bibinfo {author}
  {\bibfnamefont{C.~D.}\ \bibnamefont{Dewhurst}}, \bibinfo {author}
  {\bibfnamefont{P.~C.}\ \bibnamefont{Canfield}}, \bibinfo {author}
  {\bibfnamefont{R.~W.}\ \bibnamefont{McCallum}},\ and\ \bibinfo {author}
  {\bibfnamefont{A.~I.}\ \bibnamefont{Goldman}},\ }%
  \bibfield{journal}{%
  \bibinfo {journal} {Phys. Rev. Lett.}\ }%
  \textbf{\bibinfo {volume} {102}},\ \bibinfo {pages} {047204} (\bibinfo {year}
  {2007})%
  \bibAnnoteFile{NoStop}{kreyssig07}%
\bibitem{weissm08}%
  \BibitemOpen
  \bibfield{author}{%
  \bibinfo {author} {\bibfnamefont{G.}~\bibnamefont{Balaji}}, \bibinfo {author}
  {\bibfnamefont{S.}~\bibnamefont{Ghosh}}, \bibinfo {author}
  {\bibfnamefont{F.}~\bibnamefont{D\"obrich}}, \bibinfo {author}
  {\bibfnamefont{H.}~\bibnamefont{Eckerlebe}},\ and\ \bibinfo {author}
  {\bibfnamefont{J.}~\bibnamefont{Weissm\"uller}},\ }%
  \bibfield{journal}{%
  \bibinfo {journal} {Phys. Rev. Lett.}\ }%
  \textbf{\bibinfo {volume} {100}},\ \bibinfo {pages} {227202} (\bibinfo {year}
  {2008})%
  \bibAnnoteFile{NoStop}{weissm08}%
\bibitem{gri09}%
  \BibitemOpen
  \bibfield{author}{%
  \bibinfo {author} {\bibfnamefont{S.~V.}\ \bibnamefont{Grigoriev}}, \bibinfo
  {author} {\bibfnamefont{D.}~\bibnamefont{Chernyshov}}, \bibinfo {author}
  {\bibfnamefont{V.~A.}\ \bibnamefont{Dyadkin}}, \bibinfo {author}
  {\bibfnamefont{V.}~\bibnamefont{Dmitriev}}, \bibinfo {author}
  {\bibfnamefont{S.~V.}\ \bibnamefont{Maleyev}}, \bibinfo {author}
  {\bibfnamefont{E.~V.}\ \bibnamefont{Moskvin}}, \bibinfo {author}
  {\bibfnamefont{D.}~\bibnamefont{Menzel}}, \bibinfo {author}
  {\bibfnamefont{J.}~\bibnamefont{Schoenes}},\ and\ \bibinfo {author}
  {\bibfnamefont{H.}~\bibnamefont{Eckerlebe}},\ }%
  \bibfield{journal}{%
  \bibinfo {journal} {Phys. Rev. Lett.}\ }%
  \textbf{\bibinfo {volume} {102}},\ \bibinfo {pages} {037204} (\bibinfo {year}
  {2009})%
  \bibAnnoteFile{NoStop}{gri09}%
\bibitem{pflei2009}%
  \BibitemOpen
  \bibfield{author}{%
  \bibinfo {author} {\bibfnamefont{S.}~\bibnamefont{M\"uhlbauer}}, \bibinfo
  {author} {\bibfnamefont{B.}~\bibnamefont{Binz}}, \bibinfo {author}
  {\bibfnamefont{F.}~\bibnamefont{Jonietz}}, \bibinfo {author}
  {\bibfnamefont{C.}~\bibnamefont{Pfleiderer}}, \bibinfo {author}
  {\bibfnamefont{A.}~\bibnamefont{Rosch}}, \bibinfo {author}
  {\bibfnamefont{A.}~\bibnamefont{Neubauer}}, \bibinfo {author}
  {\bibfnamefont{R.}~\bibnamefont{Georgii}},\ and\ \bibinfo {author}
  {\bibfnamefont{P.}~\bibnamefont{B\"oni}},\ }%
  \bibfield{journal}{%
  \bibinfo {journal} {Science}\ }%
  \textbf{\bibinfo {volume} {323}},\ \bibinfo {pages} {915} (\bibinfo {year}
  {2009})%
  \bibAnnoteFile{NoStop}{pflei2009}%
\bibitem{uehland2010}%
  \BibitemOpen
  \bibfield{author}{%
  \bibinfo {author} {\bibfnamefont{B.~G.}\ \bibnamefont{Ueland}}, \bibinfo
  {author} {\bibfnamefont{J.~W.}\ \bibnamefont{Lynn}}, \bibinfo {author}
  {\bibfnamefont{M.}~\bibnamefont{Laver}}, \bibinfo {author}
  {\bibfnamefont{Y.~J.}\ \bibnamefont{Choi}},\ and\ \bibinfo {author}
  {\bibfnamefont{S.-W.}\ \bibnamefont{Cheong}},\ }%
  \bibfield{journal}{%
  \bibinfo {journal} {Phys. Rev. Lett.}\ }%
  \textbf{\bibinfo {volume} {104}},\ \bibinfo {pages} {147204} (\bibinfo {year}
  {2010})%
  \bibAnnoteFile{NoStop}{uehland2010}%
\bibitem{krycka2010}%
  \BibitemOpen
  \bibfield{author}{%
  \bibinfo {author} {\bibfnamefont{K.~L.}\ \bibnamefont{Krycka}}, \bibinfo
  {author} {\bibfnamefont{R.~A.}\ \bibnamefont{Booth}}, \bibinfo {author}
  {\bibfnamefont{C.~R.}\ \bibnamefont{Hogg}}, \bibinfo {author}
  {\bibfnamefont{Y.}~\bibnamefont{Ijiri}}, \bibinfo {author}
  {\bibfnamefont{J.~A.}\ \bibnamefont{Borchers}}, \bibinfo {author}
  {\bibfnamefont{W.~C.}\ \bibnamefont{Chen}}, \bibinfo {author}
  {\bibfnamefont{S.~M.}\ \bibnamefont{Watson}}, \bibinfo {author}
  {\bibfnamefont{M.}~\bibnamefont{Laver}}, \bibinfo {author}
  {\bibfnamefont{T.~R.}\ \bibnamefont{Gentile}}, \bibinfo {author}
  {\bibfnamefont{L.~R.}\ \bibnamefont{Dedon}}, \bibinfo {author}
  {\bibfnamefont{S.}~\bibnamefont{Harris}}, \bibinfo {author}
  {\bibfnamefont{J.~J.}\ \bibnamefont{Rhyne}},\ and\ \bibinfo {author}
  {\bibfnamefont{S.~A.}\ \bibnamefont{Majetich}},\ }%
  \bibfield{journal}{%
  \bibinfo {journal} {Phys. Rev. Lett.}\ }%
  \textbf{\bibinfo {volume} {104}},\ \bibinfo {pages} {207203} (\bibinfo {year}
  {2010})%
  \bibAnnoteFile{NoStop}{krycka2010}%
\bibitem{laver2010}%
  \BibitemOpen
  \bibfield{author}{%
  \bibinfo {author} {\bibfnamefont{M.}~\bibnamefont{Laver}}, \bibinfo {author}
  {\bibfnamefont{C.}~\bibnamefont{Mudivarthi}}, \bibinfo {author}
  {\bibfnamefont{J.~R.}\ \bibnamefont{Cullen}}, \bibinfo {author}
  {\bibfnamefont{A.~B.}\ \bibnamefont{Flatau}}, \bibinfo {author}
  {\bibfnamefont{W.-C.}\ \bibnamefont{Chen}}, \bibinfo {author}
  {\bibfnamefont{S.~M.}\ \bibnamefont{Watson}},\ and\ \bibinfo {author}
  {\bibfnamefont{M.}~\bibnamefont{Wuttig}},\ }%
  \bibfield{journal}{%
  \bibinfo {journal} {Phys. Rev. Lett.}\ }%
  \textbf{\bibinfo {volume} {105}},\ \bibinfo {pages} {027202} (\bibinfo {year}
  {2010})%
  \bibAnnoteFile{NoStop}{laver2010}%
\bibitem{michels2010epjb}%
  \BibitemOpen
  \bibfield{author}{%
  \bibinfo {author} {\bibfnamefont{D.}~\bibnamefont{Honecker}}, \bibinfo
  {author} {\bibfnamefont{A.}~\bibnamefont{Ferdinand}}, \bibinfo {author}
  {\bibfnamefont{F.}~\bibnamefont{D\"obrich}}, \bibinfo {author}
  {\bibfnamefont{C.~D.}\ \bibnamefont{Dewhurst}}, \bibinfo {author}
  {\bibfnamefont{A.}~\bibnamefont{Wiedenmann}}, \bibinfo {author}
  {\bibfnamefont{C.}~\bibnamefont{G\'{o}mez-Polo}}, \bibinfo {author}
  {\bibfnamefont{K.}~\bibnamefont{Suzuki}},\ and\ \bibinfo {author}
  {\bibfnamefont{A.}~\bibnamefont{Michels}},\ }%
  \bibfield{journal}{%
  \bibinfo {journal} {Eur. Phys. J. B}\ }%
  \textbf{\bibinfo {volume} {76}},\ \bibinfo {pages} {209} (\bibinfo {year}
  {2010})%
  \bibAnnoteFile{NoStop}{michels2010epjb}%
\bibitem{bischof07}%
  \BibitemOpen
  \bibfield{author}{%
  \bibinfo {author} {\bibfnamefont{M.}~\bibnamefont{Bischof}}, \bibinfo
  {author} {\bibfnamefont{P.}~\bibnamefont{Staron}}, \bibinfo {author}
  {\bibfnamefont{A.}~\bibnamefont{Michels}}, \bibinfo {author}
  {\bibfnamefont{P.}~\bibnamefont{Granitzer}}, \bibinfo {author}
  {\bibfnamefont{K.}~\bibnamefont{Rumpf}}, \bibinfo {author}
  {\bibfnamefont{H.}~\bibnamefont{Leitner}}, \bibinfo {author}
  {\bibfnamefont{C.}~\bibnamefont{Scheu}},\ and\ \bibinfo {author}
  {\bibfnamefont{H.}~\bibnamefont{Clemens}},\ }%
  \bibfield{journal}{%
  \bibinfo {journal} {Acta mater.}\ }%
  \textbf{\bibinfo {volume} {55}},\ \bibinfo {pages} {2637} (\bibinfo {year}
  {2007})%
  \bibAnnoteFile{NoStop}{bischof07}%
\bibitem{micmagref}%
  \BibitemOpen
  \bibinfo {note} {See, e.g., {\it Handbook of Magnetism and Advanced Magnetic
  Materials}, edited by H. Kronm\"uller and S. Parkin, Vol.~2: {\it
  Micromagnetism} (Wiley, Chichester, 2007).}%
  \bibAnnoteFile{Stop}{micmagref}%
\bibitem{dufour2011}%
  \BibitemOpen
  \bibfield{author}{%
  \bibinfo {author} {\bibfnamefont{C.}~\bibnamefont{Dufour}}, \bibinfo {author}
  {\bibfnamefont{M.~R.}\ \bibnamefont{Fitzsimmons}}, \bibinfo {author}
  {\bibfnamefont{J.~A.}\ \bibnamefont{Borchers}}, \bibinfo {author}
  {\bibfnamefont{M.}~\bibnamefont{Laver}}, \bibinfo {author}
  {\bibfnamefont{K.~L.}\ \bibnamefont{Krycka}}, \bibinfo {author}
  {\bibfnamefont{K.}~\bibnamefont{Dumesnil}}, \bibinfo {author}
  {\bibfnamefont{S.~M.}\ \bibnamefont{Watson}}, \bibinfo {author}
  {\bibfnamefont{W.~C.}\ \bibnamefont{Chen}}, \bibinfo {author}
  {\bibfnamefont{J.}~\bibnamefont{Won}},\ and\ \bibinfo {author}
  {\bibfnamefont{S.}~\bibnamefont{Singh}},\ }%
  \bibfield{journal}{%
  \bibinfo {journal} {Phys. Rev. B}\ }%
  \textbf{\bibinfo {volume} {84}},\ \bibinfo {pages} {064420} (\bibinfo {year}
  {2011})%
  \bibAnnoteFile{NoStop}{dufour2011}%
\bibitem{michels06prb}%
  \BibitemOpen
  \bibfield{author}{%
  \bibinfo {author} {\bibfnamefont{A.}~\bibnamefont{Michels}}, \bibinfo
  {author} {\bibfnamefont{C.}~\bibnamefont{Vecchini}}, \bibinfo {author}
  {\bibfnamefont{O.}~\bibnamefont{Moze}}, \bibinfo {author}
  {\bibfnamefont{K.}~\bibnamefont{Suzuki}}, \bibinfo {author}
  {\bibfnamefont{P.~K.}\ \bibnamefont{Pranzas}}, \bibinfo {author}
  {\bibfnamefont{J.}~\bibnamefont{Kohlbrecher}},\ and\ \bibinfo {author}
  {\bibfnamefont{J.}~\bibnamefont{Weissm\"uller}},\ }%
  \bibfield{journal}{%
  \bibinfo {journal} {Phys. Rev. B}\ }%
  \textbf{\bibinfo {volume} {74}},\ \bibinfo {pages} {134407} (\bibinfo {year}
  {2006})%
  \bibAnnoteFile{NoStop}{michels06prb}%
\bibitem{erokhin2011ieee}%
  \BibitemOpen
  \bibfield{author}{%
  \bibinfo {author} {\bibfnamefont{S.}~\bibnamefont{Erokhin}}, \bibinfo
  {author} {\bibfnamefont{D.}~\bibnamefont{Berkov}}, \bibinfo {author}
  {\bibfnamefont{N.}~\bibnamefont{Gorn}},\ and\ \bibinfo {author}
  {\bibfnamefont{A.}~\bibnamefont{Michels}},\ }%
  \bibfield{journal}{%
  \bibinfo {journal} {IEEE Trans. Magn.}\ }%
  \textbf{\bibinfo {volume} {47}},\ \bibinfo {pages} {3044} (\bibinfo {year}
  {2011}),\ \bibinfo {note} {and S. Erokhin, D. Berkov, N. Gorn, and A.
  Michels, Phys. Rev. B, submitted.}%
  \bibAnnoteFile{Stop}{erokhin2011ieee}%
\bibitem{michels08epl}%
  \BibitemOpen
  \bibfield{author}{%
  \bibinfo {author} {\bibfnamefont{A.}~\bibnamefont{Michels}}, \bibinfo
  {author} {\bibfnamefont{F.}~\bibnamefont{D\"obrich}}, \bibinfo {author}
  {\bibfnamefont{M.}~\bibnamefont{Elmas}}, \bibinfo {author}
  {\bibfnamefont{A.}~\bibnamefont{Ferdinand}}, \bibinfo {author}
  {\bibfnamefont{J.}~\bibnamefont{Markmann}}, \bibinfo {author}
  {\bibfnamefont{M.}~\bibnamefont{Sharp}}, \bibinfo {author}
  {\bibfnamefont{H.}~\bibnamefont{Eckerlebe}}, \bibinfo {author}
  {\bibfnamefont{J.}~\bibnamefont{Kohlbrecher}},\ and\ \bibinfo {author}
  {\bibfnamefont{R.}~\bibnamefont{Birringer}},\ }%
  \bibfield{journal}{%
  \bibinfo {journal} {EPL}\ }%
  \textbf{\bibinfo {volume} {81}},\ \bibinfo {pages} {66003} (\bibinfo {year}
  {2008})%
  \bibAnnoteFile{NoStop}{michels08epl}%
\bibitem{elmas09}%
  \BibitemOpen
  \bibfield{author}{%
  \bibinfo {author} {\bibfnamefont{A.}~\bibnamefont{Michels}}, \bibinfo
  {author} {\bibfnamefont{M.}~\bibnamefont{Elmas}}, \bibinfo {author}
  {\bibfnamefont{F.}~\bibnamefont{D\"obrich}}, \bibinfo {author}
  {\bibfnamefont{M.}~\bibnamefont{Ames}}, \bibinfo {author}
  {\bibfnamefont{J.}~\bibnamefont{Markmann}}, \bibinfo {author}
  {\bibfnamefont{M.}~\bibnamefont{Sharp}}, \bibinfo {author}
  {\bibfnamefont{H.}~\bibnamefont{Eckerlebe}}, \bibinfo {author}
  {\bibfnamefont{J.}~\bibnamefont{Kohlbrecher}},\ and\ \bibinfo {author}
  {\bibfnamefont{R.}~\bibnamefont{Birringer}},\ }%
  \bibfield{journal}{%
  \bibinfo {journal} {EPL}\ }%
  \textbf{\bibinfo {volume} {85}},\ \bibinfo {pages} {47003} (\bibinfo {year}
  {2009})%
  \bibAnnoteFile{NoStop}{elmas09}%
\bibitem{commentprl2012}%
  \BibitemOpen
  \bibinfo {note} {When the dipolar interaction is ignored in the micromagnetic
  computations (irrespective of the value of $\Delta M$), all Fourier
  coefficients are isotropic for all applied fields investigated (data not
  shown). This underlines the importance of dipolar correlations for the
  magnetic microstructure and associated SANS of nanomagnets.}%
  \bibAnnoteFile{Stop}{commentprl2012}%
\end{thebibliography}
\end{document}